# Identification of single nucleotides in MoS$_2$ nanopores

Jiandong Feng[1#], Ke Liu[1#], Roman D. Bulushev[1], Sergey Khlybov[1], Dumitru Dumcenco [2], Andras. Kis[2], Aleksandra Radenovic[1*]

[1] *Laboratory of Nanoscale Biology, Institute of Bioengineering, School of Engineering, EPFL, 1015 Lausanne, Switzerland*

[2] *Laboratory of Nanoscale Electronics and Structure, Institute of Electrical Engineering , School of Engineering, EPFL, 1015 Lausanne, Switzerland*

*correspondence should be addressed to aleksandra.radenovic@epfl.ch
[#] equal contribution

**ABSTRACT**

**Ultrathin membranes have drawn much attention due to their unprecedented spatial resolution for DNA nanopore sequencing. However, the high translocation velocity (3000-50000 nt/ms) of DNA molecules moving across such membranes limits their usability. To this end, we have introduced a viscosity gradient system based on room-temperature ionic liquids (RTILs) to control the dynamics of DNA translocation through a nanometer-size pore fabricated in an atomically thin MoS$_2$ membrane. This allows us for the first time to statistically identify all four types of nucleotides with solid state nanopores. Nucleotides are identified according to the current signatures recorded during their transient residence in the narrow orifice of the atomically thin MoS$_2$ nanopore. In this novel architecture that exploits high viscosity of RTIL, we demonstrate single-**

**nucleotide translocation velocity that is an optimal speed (1-50 nt/ms) for DNA sequencing, while keeping the signal to noise ratio (SNR) higher than 10. Our findings pave the way for future low-cost and rapid DNA sequencing using solid-state nanopores.**

**INTRODUCTION**

Translocation velocity of DNA molecules in solid-state nanopores is on the order of 3000-50000 nt/ms[1]. This large translocation velocity range originates from different parameters such as the wide range of the pore sizes (1.5-25 nm) and applied potentials (100 mV-800 mV)[1].

The high translocation velocity of DNA molecules, together with a low ionic current signal-to-noise ratio and a relatively large sensing region due to the pore membrane thickness, that is typically 10 -20 nm and therefore can accommodate 30-60 nucleotides at a time[1], has been a major obstacle for achieving sequencing data in solid state nanopores. Although single nucleotide identification[2,3] and DNA sequencing using biological pores have already been demonstrated[3,4] their fragility, difficulties related to measuring pA-range ionic currents together with their dependence on biochemical reagents, make solid state nanopores an attractive alternative[5]. In contrast to bio-engineered pores, solid state nanopores can operate in various liquid media and pH conditions, their production is scalable and compatible with nanofabrication techniques and does not require the excessive use of biochemical reagents. All these advantages are expected to lower the cost of sequencing. The basic sensing principle is the same as in bio-engineered pores. Ideally, the sequence of nucleotides, genetic information, along a single DNA molecule can be registered by monitoring small changes in the ionic current caused by the transient residing of single nucleotides within a nanometer-size pore[6]. In addition, solid state nanopores allow a transverse detection scheme, based on detecting changes in the electrical conductivity of a thin semiconducting channel caused by the translocating molecule. The DNA translocations in biological nanopores are currently too slow, on the other hand in solid-state nanopores are too fast compared to the optimal DNA sequencing velocity of 1-50 nt/ms [7].

Achieving optimal translocation speed for both biological and solid state nanopores remains a significant challenge[8-18]. Here, we demonstrate that molecular translocation speeds in a nanopore sensing system can be decreased by two to three orders of magnitude using an ionic liquid/water viscosity gradient system together with a nanopore fabricated in atomically thin membranes of molybdenum disulfide ($MoS_2$) which fulfil the requirement for spatial resolution. This requirement has been first met with the graphene membranes [19-21]. By using graphene, the thinnest known material, the ultimate end thickness of the membrane for solid state nanopores has been reached[19-21]. Nanopores realized in three-layer graphite structures having thickness of 1 nm should display higher signal to noise ratio compared to single –layer graphene nanopores[22,23]. The use of 2D materials such as graphene is particularly interesting since it allows concomitant detection of DNA translocation with two synchronized signals, i.e., ionic current in the nanopore and the current in the graphene nanoribbons (GNR) as recently demonstrated by our group [5]. However, pristine graphene nanopores exhibit strong hydrophobic interactions with DNA [24] that limit their long-term use due to clogging, requiring the use of surface functionalization[25,26]. In parallel, other 2D materials, such as boron nitride (BN)[26] and $MoS_2$[27] have been implemented as alternatives to graphene while fabrication advances allowed for realization of nanopores in ultrathin $SiN_x$[28] and $HfO_2$[18]. $MoS_2$ nanopores are particularly interesting since they can be used for extended periods of time (hours even days) without any additional functionalization[27]. Sticking behaviour of DNA to $MoS_2$ nanopore is reduced due to the Mo-rich region around the drilled pore after TEM irradiation[29] while the stability could be attributed to its thickness.

Single-layer $MoS_2$ has a thickness of 0.7 nm and a direct band gap of at least 1.8 eV[5,30] which is essential for electronic base detection in field-effect transistors (FETs) [5,31] thus making $MoS_2$ a promising material for single-nucleotide detection, as recently computationally demonstrated by Farimani *et.al.* [32]

Our approach to slow down DNA translocation, has been inspired by the remarkable physical and chemical properties of RTILs, non-aqueous electrolytes composed of a pair of organic cations and anions. RTILs have been termed "magical chemicals", due to the high degree of freedom in fine-tuning their structure that allows tailoring physical and chemical properties for a given application[33]. We have chosen 1-butyl-3-methylimidazolium hexafluorophosphate ($BmimPF_6$), since it has a broad viscosity window of 10-400 cP[34], which can be tuned to optimize the temporal resolution. Tunability can be obtained either by varying the temperature (20°C-50 °C) or by mixing $BmimPF_6$ and $BmimPF_4$ in different ratios[35]. $BmimPF_6$ is also a friendly solvent for bio-molecules and most importantly, it exhibits good ionic conductivity of 1.4 mS cm$^{-1}$ [34]. It has also been widely used as an electrolyte with a wide electrochemical window[36]. In contrast, the low conductivity of glycerol limited previous attempts to a narrow viscosity window (1.2-5 cP) and consequently achieved only modest improvement in DNA translocation speed (3000 nt/ms)[8]. In our viscosity gradient system, schematically shown on **Fig. 1a,** it was possible to employ pure BmimPF$_6$ without compromising the conductance of the MoS$_2$ nanopore for all tested nanopore devices listed in SI Table 2. Details on the properties and fabrication of MoS$_2$ nanopores can be found in Methods section of this paper and in Liu *et al.*[27]

For example even in the large nanopore 17 ± 2 nm conductance in pure RTILs is relatively low (~1 nS)[37] when compared to KCl, **Fig. 1c**, inspired by the use of concentration gradient systems in nanopores[38], we have realized a viscosity and concentration gradient system with the conductivity of 210 nS. The *cis* chamber in our system contains RTILs (BmimPF$_6$) while *trans* chamber contains 2M aqueous KCl solution. It is important to note that here we use two types of solvents with completely different physicochemical properties and that in the region inside and close to the pore we have in fact a non-homogeneous phase solution. The conductance of 280 nS measured in 2M KCl/2M KCl is reduced to 210 nS in our viscosity gradient system as shown in **Fig. 1c**. To gain insight into the ionic transport through the nanopores in the presence of an inhomogeneous phase solution, we have performed finite element analysis by solving the Poisson–Nernst–Planck (PNP) equation. **Fig 1.d.** shows the mass fraction of water molecules, anions and cations as a function of distance from the nanopore at a transmembrane bias voltage of 400 mV. The sub-nanometer membrane thickness ensured that a relatively high number of water molecules diffused from the *trans* into the *cis* chamber. Similarly, anions and cations diffused into their respective chambers. Modeled conductances for 2.5 nm, 5 nm 10 nm pore, shown in **SI Fig. 1.d-f,** are in good agreement with our measurements (**SI Fig. 1.a-c**). Interestingly, the mass fraction of water molecules in the *cis* chamber shows a weak dependence on the transmembrane bias, while $PF_6^-$ diffusion is strongly affected (**SI Fig. 1.g**). Having successfully built and characterized our viscosity gradient system, we performed our first translocation experiment by adding 48.5 kbp λ-dsDNA to the *cis* chamber filled with $BmimPF_6$. In order to minimize the contribution from the nanopore-DNA interaction that can also significantly slow down DNA translocation[28], we decided to first use $MoS_2$ nanopores with relatively large diameters (~20 nm, **SI Fig. 2.a.**). Fig. 2a displays the typical current trace recorded during the translocation of the λ-DNA molecule in the viscosity gradient system in the presence of a transmembrane bias voltage of 400 mV. When compared to a typical current trace acquired in a 2M aqueous KCl solution obtained using the same pore and transmembrane voltage, one can observe temporal improvement and no reduction in the amplitude of the current drop. Unlike other viscous systems for slowing down DNA translocation, signal amplitude has been preserved owing to the conductive nature of RTILs and high concentration of chloride ions inside the pore. The average translocation time is 130 ms for λ-DNA in the viscosity gradient system, and 1.4 ms in the 2M KCl solution presenting two orders of magnitude's improvement.

At this point, in the absence of the electro-osmotic flow (EOF) and charge reduction for a given pore, DNA molecule and bias voltage, we can introduce the retardation factor *r* (for details see **SI**). We obtain a retardation factor higher than 100 that is predominantly due to the increase in the viscosity in our viscosity gradient system. However, scatter plots and DNA translocation histograms in **Fig. 2b and Fig. 2c** reveal a large spread in dwell times that can be attributed to several factors associated with the viscosity gradient system. In reality, EOF, charge reduction as well as long-range hydrodynamic effects and the existence of gradients in the free-energy landscape have to be included in the future model and could result in a more complex dynamics of DNA translocation in the viscosity gradient system than assumed in our simplistic model (presented in SI). In addition, it is possible that we have overestimated the value of $BmimPF_6$ viscosity in the vicinity of the pore. More accurate calculation of the retardation factor should include the effects related to charge reduction and the presence of the EOF[39]. Due to the negative charges at the surface of $MoS_2$ membrane and within the pore, the direction of EOF is opposite to the direction of DNA translocations and could result in further slowing down. By comparing translocation traces before, during and after translocation events we see that they all have a similar noise level of 520-540 pA (**SI Fig. 3**.). We observe a slight increase of noise during the translocation that can be explained by the fact that DNA interacts strongly with $BmimPF_6$ via electrostatic interaction between the cationic $Bmim^+$ groups and DNA phosphates (P-O bonds)[40]. Because of this electrostatic interaction and the hydrophobic association between Bmim$^+$ and bases, DNA molecules can act as carriers for Bmim$^+$ ions from the *cis* to the *trans* chamber.

In general, the single-molecule DNA translocation process can be viewed as a voltage-driven barrier crossing as shown in **SI Fig. 4 a**. To further explain retardation mechanism, we explore the voltage dependence of pNEB 193 (2700 bp long DNA plasmid) translocation dwell times in the MoS$_2$ nanopore. The observed power law scaling is consistent with Kramer′s theory **SI Fig. 4 a, b**. A free-energy barrier predominately arises from the RTILs and KCl/water interface and includes a change in conformational entropy of the translocating polymer. The threading process across nanopore in a high-voltage regime follows a force balance model detailed in **Supplementary Material.** pNEB is almost 18 times shorter than λ -DNA, however we still observe large retardation when comparing average dwell times recorded at 400 mV, under the condition of viscosity gradient 2±0.5 ms, and 40±10 μs in the 2M KCl aqueous solution.

To exploit the full potential of our viscosity gradient system, we translocate short homopolymers, poly(dA)$_{30}$, poly(dT)$_{30}$, poly(dG)$_{30}$ and poly(dC)$_{30}$, through a 2.8 nm diameter pore in single-layer MoS$_2$, shown in the TEM micrograph on **SI Fig.2b**. DNA-pore interactions can also increase the translocation time by one order of magnitude[12]. The 2.8 nm pore in single layer MoS$_2$ membrane suspended over smaller opening in the nitride, even without any special pretreatment[41], displays better noise properties with a current RMS of 59 pA at 0mV and 89 pA at 200 mV (as shown in **SI Fig. 7.** ) compared to the pores suspended over larger openings. The noise reduction is achieved by restricting the opening for freestanding MoS$_2$ membrane to the hole having 100 nm diameter. [42] Self-organization of certain ionic liquids can be further exploited to reduce 1/f noise in single nanopores as shown by Tessarit *et.al.*[43] **Fig.3** shows translocation traces of short DNA homopolymers for periods of 0.5 s and 0.1 s respectively. Four peaks can be clearly distinguished in the histogram of current drops shown in **Fig. 3b.** The density scatter plots shown in **SI Fig. 8a** are useful in revealing the range of the most probable dwell time for the four types of poly-nucleotides at the transmembrane bias voltage of 200 mV. The current traces and histogram of poly(dG)$_{30}$ homopolymer display two peaks. However, from the poly(dG)$_{30}$ density scatter plot (**SI Fig 8**) one can easily identify which peak is more probable. Based on the amplitude and temporal signature of the second peak, we believe that it might originate from the G-quadruplex formation [44]. Venta *et al.* [28] reported a much faster translocation (20 µs) of such homoplymers using a 1M Hz amplifier in 1.5 nm pores with high applied voltage (1V). However, high bandwidth amplifier introduced additional noise and high voltage might reduce the lifetime of the device. In the viscosity gradient system and 2.8 nm pore, we achieved 10-50 times slowing down compared to the results from Venta *et al.* [28]

    Finally, using the same 2.8 nm diameter MoS$_2$ nanopore, we translocate single nucleotides, dAMP, dTMP, dGMP and dCMP. The exceptional durability of the MoS$_2$ nanopore has allowed us to perform 8 consecutive experiments with high

throughput (more than 10000 events collected, enabling robust statistical analysis) (**Fig. 3** and **Fig. 4**) in the same pore. Each experiment has been preceded with the flushing of the fluidics and with the short control experiment to confirm the absence of the analyte from the previous experiment. Not only does this show the extraordinary resilience of our nanopores, but, to our surprise, the dwell times of single nucleotides are comparable to those of 30mer homopolymers. At this scales, when comparing the dwell times of single nucleotides to homopolymers, one needs to account for the charge reduction difference that will result in the lower net force acting on the single nucleotide compared to the homopolymers. In the pores with diameters < 5nm, observed translocation retardation is a cumulative effect that includes several components: interaction of the translocating molecule with the pore wall, electrostatic interaction between $Bmim^+$ cations and phosphate groups of DNA, the hydrophobic association between $Bmim^+$ and DNA bases[45] and finally, the viscosity gradient. The contribution of the viscosity gradient to the retardation will increase with the increasing DNA length. Consequently, for single nucleotides this contribution is decreased, however due to the charge reduction, the lower net force acting on the single nucleotide might account for observed long translocation times of single nucleotides.

The use of single-layer $MoS_2$ as the membrane material and the viscosity gradient system in combination with the small nanopore have been crucial for the single nucleotide discrimination as shown in **Fig. 4. SI Fig.9** shows translocation traces of four single nucleotides for periods of 0.5 s and 0.1 s respectively. Here, the obtained translocation speed is in the range from 1-50 nt/ms. In accordance with the physical dimensions for four nucleotides, we observe for dGMP, centered at 0.8 nA and a smallest current drop for the smallest single nucleotide dCMP, centered at 0.3 nA. These observations are in good agreement with the results obtained on single nucleotide discrimination using protein pores [2-4]. Although the current drop for dAMP is slightly larger than dTMP (0.65 nA compared to 0.45 nA), we believe that this inconsistency might be due to the stronger Bmim$^+$ affinity towards dAMP compared to dTMP [46]. It has been established that RTILs could selectively bind to DNA[40], while RTILs based on metal chelate anions could be designed to have specific bonding to the bases[47]. In our system, this could be further exploited to amplify the small differences in bases. Using only ionic current drops of 500-3000 events for four nucleotides, we performed a Welch's t-test and found p-values to be all less than 0.0001. Moreover, this simple statistical analysis revealed a minimum event number to be 6-9 for nucleotide identification with a confidence of 99%. With the addition of the other parameters such as dwell time it might be possible to identify single nucleotides with one read while the presence of a direct band gap in MoS$_2$ should allow for straightforward multiplexing in a detection scheme based on the transverse current. We have also reproduced the discrimination of single nucleotides in a slightly bigger pore with a diameter of 3.3 nm under the same conditions of the viscosity gradient system (**SI Fig. 10**) and with a similar number of events (>10000).

When translocating single nucleotides (such as dAMP) and increasing the potential (from 200 mV to 400 mV), we have observed increases in the current drop amplitudes as shown on **SI Fig. 11**, further confirming that observed events are indeed from single nucleotide translocation. However, applying a higher potential caused an increase in the noise. The data presented on **SI Fig. 10** and **SI Fig. 12** are therefore collected at 200 mV. As expected, in the 3.3 nm pore, the translocation events (>10000) were faster and produced smaller current amplitude drops compared to the 2.8 nm pore and other smaller pores (**SI Fig. 12**). However, the trend of current drops for different types of nucleotides remained the same, as shown in **Fig 4.** (dGMP>dAMP>dTMP>dCMP). Similarly as we did for the 2.8 nm pore, by performing the Welch's t-test, we found that 14 events are needed for nucleotide identification with 99% confidence. **SI Fig. 12.** shows correlation between mean current drops related to four nucleotides and pore sizes. The dashed line placed between 3.5 nm and 4 nm indicates the maximum pore size that still allows nucleotide differentiation. In addition, translocating nucleotides in pores as small as 2 ± 0.2 nm can dramatically increase SNR up to 16.

To conclude, we have demonstrated that single-nucleotide identification can be achieved in $MoS_2$ nanopores by using a viscosity gradient to regulate the translocation speed. The viscosity gradient system can not only be used in standard ionic sensing experiments but it can be potentially combined with other schemes of nanopore sensing such as transverse current signal detection. The ultrahigh viscosity of ionic liquids results in reduced capture rates. Therefore, an optimal experimental configuration would capitalize on high-end electronics[28] and the viscosity gradient system presented here with a suitable capture rate. We believe that combining ionic liquids and monolayer $MoS_2$ nanopores, together with the readout of transverse current, either using the tunneling[48,49] or FET modality[5,31], would reach all the necessary requirements for DNA strand-sequencing such as the optimal time resolution and signal resolution in a platform that allows multiplexing, thus reducing the costs and enhancing the signal statistics.

**Acknowledgements**

We thank Prof. Jianmin Wu and Dr. Haijuan Zhang at Institute of Microanalytical System, Department of Chemistry, Zhejiang University, China for discussion in physicochemical characteristics of ionic liquids. We thank Prof. Paolo De los Rios for discussion on the force drag mechanism. We thank the Centre Interdisciplinaire de Microscopie Electronique (CIME) at EPFL for access to electron microscopes; special thanks to D. T .L. Alexander for providing training and technical assistance with TEM. Devices fabrication was partially carried out at the EPFL Center for Micro/Nanotechnology (CMi). K. L. thanks Mr. Oriol Lopez Sanchez for the training. We thank all the members from LBEN and LANES for assistance and discussion. This work was financially supported by European Research Council (grant no. 259398, PorABEL: Nanopore integrated nanoelectrodes for biomolecular manipulation and sensing and SNF Sinergia grant no. 147607). We thank Prof. Cees Dekker group and reviewers for carefully reading our manuscript and for giving detailed comments and suggestions that have been helpful to improve the manuscript.

**Author contributions**

J.F., K.L. and A.R designed research. J.F., K.L. fabricated and characterized devices, performed experiments and analyzed data. R.D.B. performed single molecule fluorescence measurements, dynamic light scattering measurements and optical trapping experiments. S.K and R.D.B. performed COMSOL modeling. D. D. performed CVD $MoS_2$ growth with A.K's supervision. J.F., K.L. A.K. and A.R. wrote the paper. All authors provided important suggestions for the experiments, discussed the results and contributed to the manuscript.

**FIGURES**

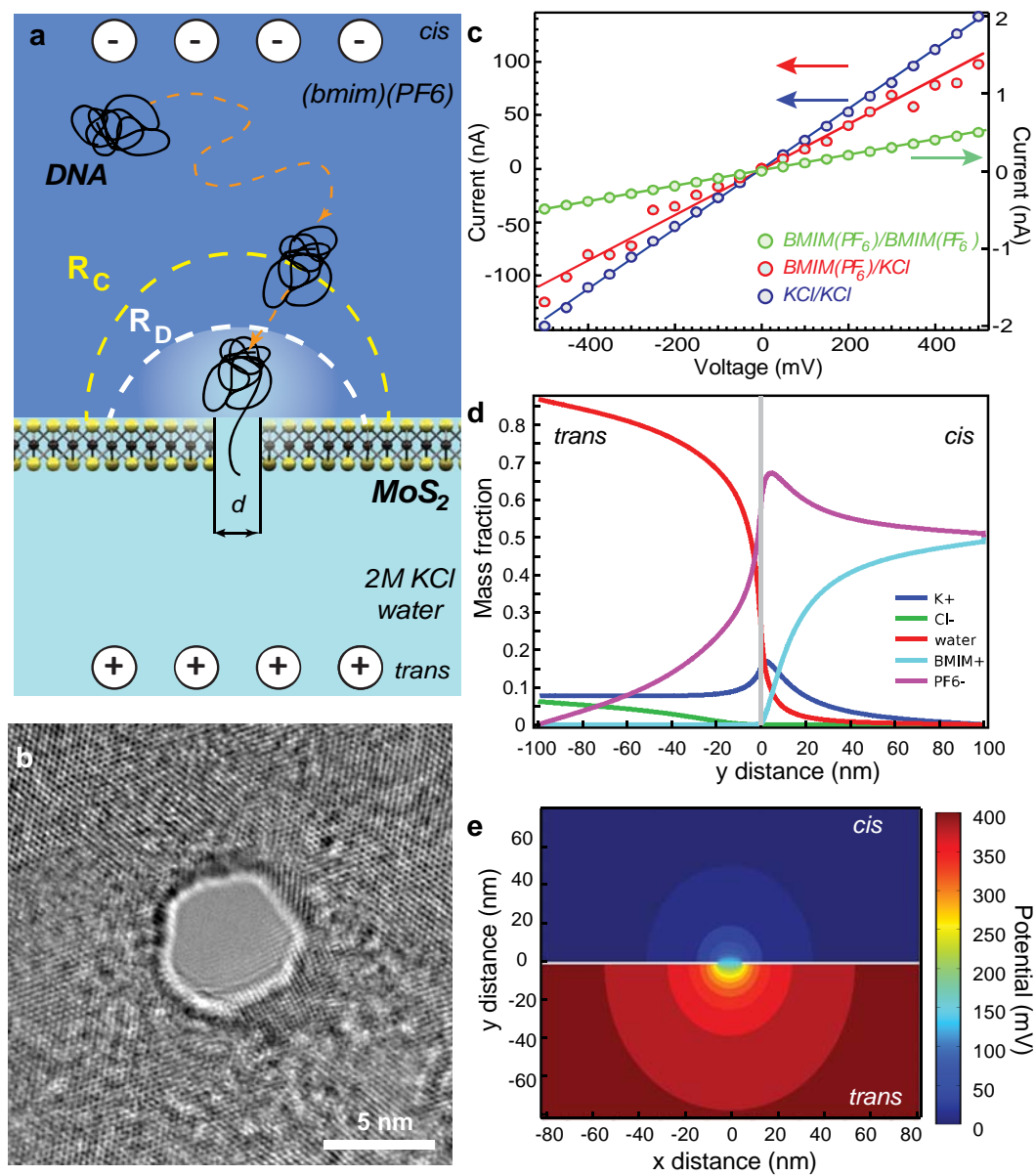

**Fig. 1**. **Schematic and characterization of the RTILs/ KCl viscosity gradient system in a nanopores based on a 2d material. (a)** *Cis* chamber contains RTILs (BmimPF$_6$) while *trans* chamber contains 2M aqueous KCl solution. The two chambers are separated by a monolayer MoS$_2$ membrane with a nanopore. Schematic also displays dynamics of DNA translocation through a monolayer MoS$_2$ nanopore. Away from the pore, DNA motion is purely diffusive due to the negligible electric field, but once within the area of capture radius Rc, DNA will be accelerated towards the pore by the force due to electrophoretic and electroosmotic effects. A part of DNA will undergo conformational change and one end will dive into the pore. The non-translocated part of the DNA polymer -monomers will keep the coil conformation and experience a strong Stokes dragging force from the ionic liquids. Consequently, DNA translocation

through the pore can be significantly slowed down. **(b)** Bright-field TEM image of a 5-nm solid-state pore fabricated in a monolayer MoS$_2$ membrane suspended over a 200 nm x 200 nm etched area formed in the center of a 20 μm large low-stress SiN$_x$ membrane with a thickness of being 20 nm. **(c)** Ohmic current–voltage responses of a 17 ± 2 nm MoS$_2$ pore. IV characteristics are taken at room temperature in a 2M aqueous KCl solution (blue circles), pure BmimPF$_6$ (green circles) and in BmimPF$_6$/2M KCl gradient (red circles) **(d)** Mass fraction of water, anions (PF$_6^-$ and Cl$^-$), cations (Bmim$^+$ and K$^+$) as a function of distance from the nanopore (note that the calculation has been performed at -400 mV) **(e)** Electric potential map evaluated numerically for the viscosity gradient system shown in (a).

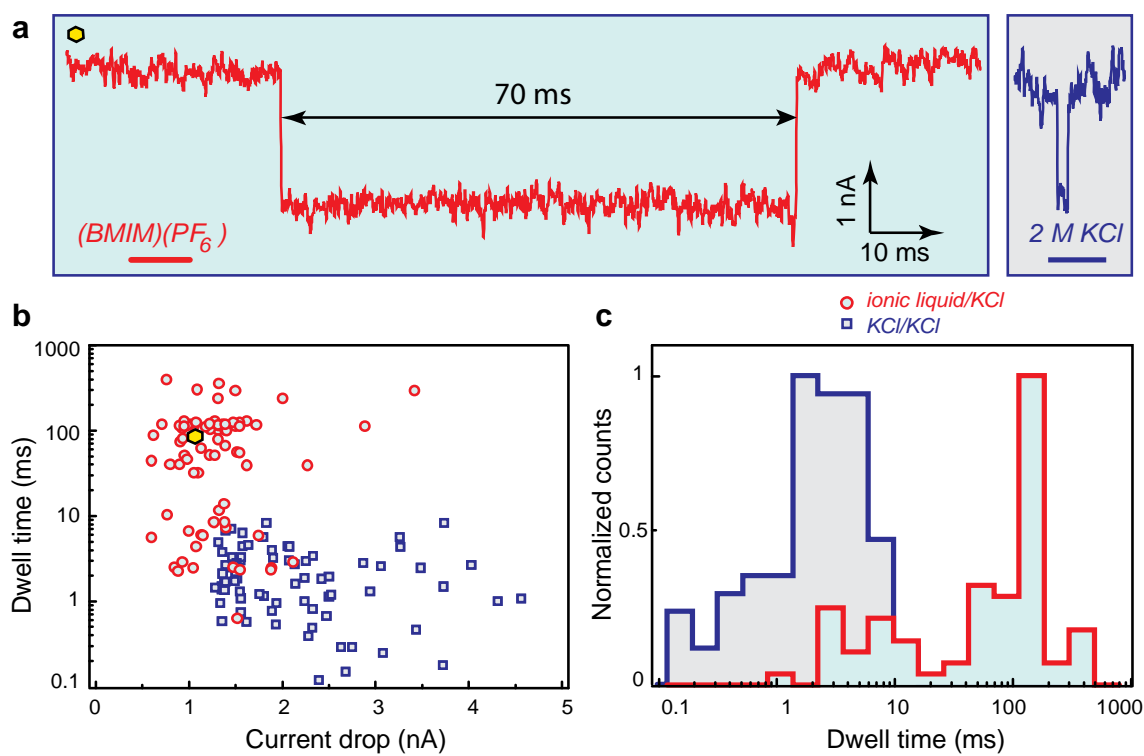

**Fig. 2. Slowing down DNA translocation by increasing the electroosmotic Stokes force $F_S$ in monolayer MoS$_2$ nanopore.** (a) An example of a 48.5 kbp λ-dsDNA translocation event in a viscosity gradient system. The corresponding current drop represents a single DNA molecule passing through the MoS$_2$ pore with a diameter of 20nm. On the right, we show a typical translocation trace for 48.5 kbp λ-dsDNA obtained using the same nanopore in the absence of the viscosity gradient, resulting in translocation times that are two orders of magnitude shorter. (Displayed traces down-sampled to 10 kHz) (b) Scatter plots (blockade current versus dwell time) for dwell time versus current signal of λ-dsDNA translocation in water (blue squares), and in our viscosity gradient system (red circles) obtained using the same 20 nm diameter MoS$_2$ nanopore. Yellow hexagon indicates the position of the event shown in (a) in respect to other events displayed in (b) (c) Histograms of translocation times corresponding to the translocation of λ-dsDNA in water (blue) and the viscosity gradient system (red).

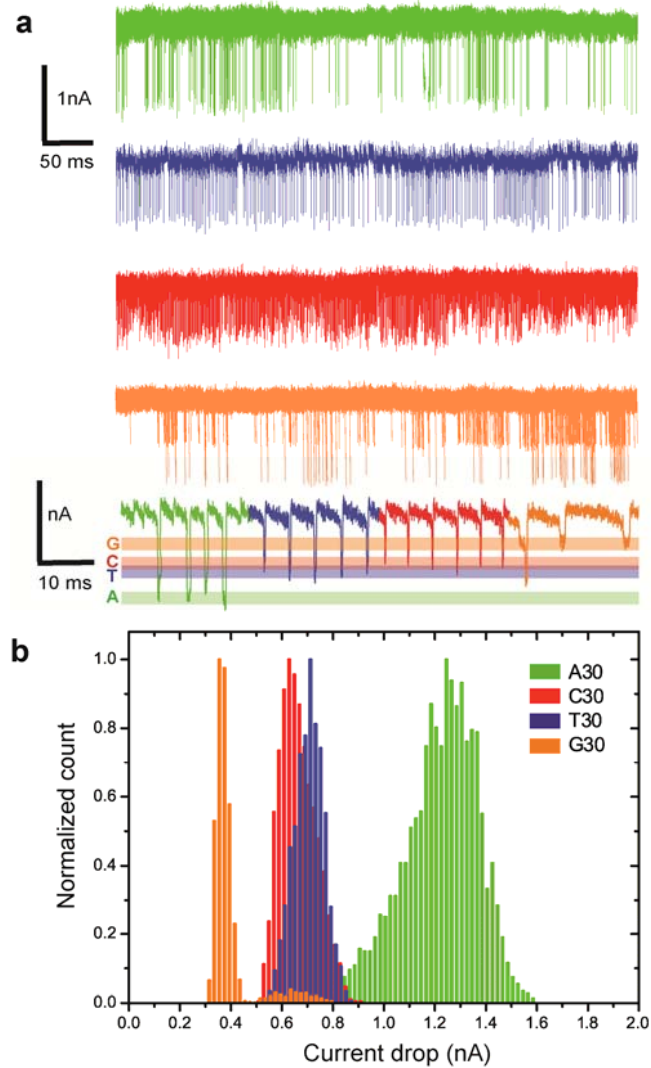

**Fig. 3**. **Identification of four 30mer oligonucleotides in the MoS$_2$ nanopore.** (a) 0.5 s and 0.1 s translocation signals for each homopolymer, poly A30 (green), poly C30 (red), poly T30 (blue) and poly G30 (orange). (b) Normalized histogram of current drops for each kind of the DNA homopolymer. The mean value for poly A30 is 1.25±0.12 nA, for poly C30 is 0.64±0.07 nA, for poly T30 is 0.71±0.06 nA and for poly G30 is 0.36±0.03 nA. Data acquired in pure RTIL cis chamber, 100 mM KCl, 25 mM Tris HCl, pH 7.5, trans chamber, at +200 mV. The concentration of short DNA homopolymers in RTILs is 0.02 μmol/ml.

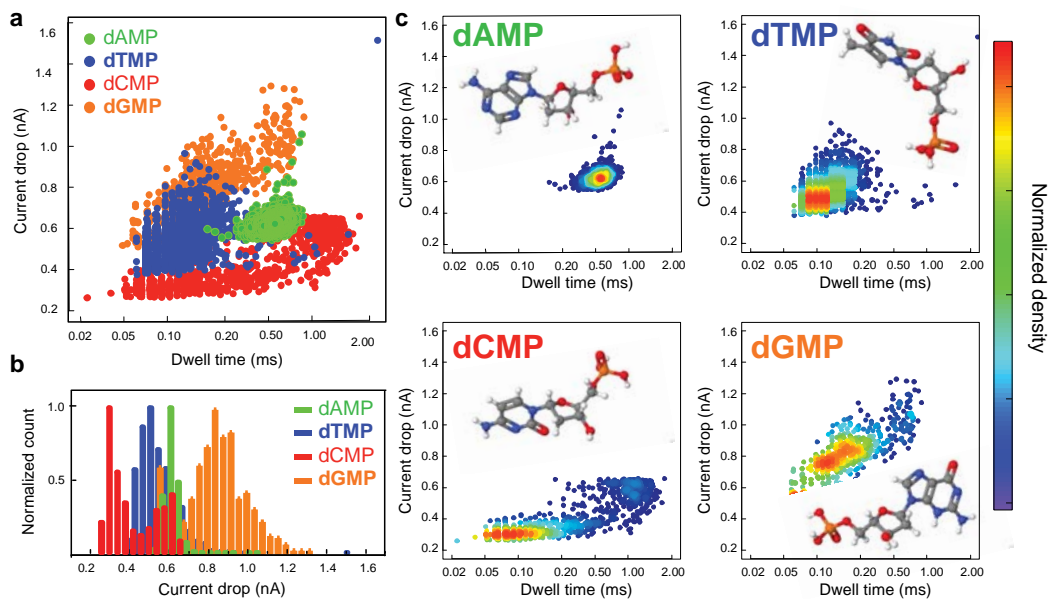

**Fig. 4**. **Identification of single nucleotides in a MoS$_2$ nanopore.** (**a**) Scatter plots of nucleotide translocation events, showing distinguished current drops and dwell times for dAMP (green), dCMP (red), dTMP (blue), and dGMP (orange). (**b**) Normalized histogram of current drops for dAMP, dTMP, dCMP, dGMP. (**c**) Density plot of single nucleotides in MoS$_2$ nanopore; for dAMP, the position of the hot spot is (0.5, 0.62), for dTMP, (0.09, 0.49), for dCMP, (0.06. 0.31) and for dGMP (0.15, 0.83). The color-map at the right shows the normalized density distribution of events. Data acquired in pure RTIL cis chamber, 100 mM KCl, 25 mM Tris HCl, pH 7.5, trans chamber, at +200 mV. The nucleotide concentration in RTILs was 5ug/ml. Insets show 3D models for the chemical structure of nucleotides.

# Supporting Material

# for

# Identification of single nucleotides in MoS$_2$ nanopores

Jiandong Feng[1#], Ke Liu[1#], Roman D. Bulushev[1], Sergey Khlybov[1], Dumitru Dumcenco[2], Andras Kis[2], Aleksandra Radenovic[1*]

[1] *Laboratory of Nanoscale Biology, Institute of Bioengineering, School of Engineering, EPFL, 1015 Lausanne, Switzerland*

[2] *Laboratory of Nanoscale Electronics and Structure, Institute of Electrical Engineering , School of Engineering, EPFL, 1015 Lausanne, Switzerland*

*\*correspondence should be addressed to [aleksandra.radenovic@epfl.ch](aleksandra.radenovic@epfl.ch)*
*[#] equal contribution*

# Setup

We fabricated devices using the previously published procedure[1]. Briefly, exfoliated or CVD-grown[2,3] thin layers of $MoS_2$ were transferred either from $SiO_2$ or sapphire substrates and suspended on $SiN_x$ membranes. Nanopores were further drilled using a JEOL 2200FS high-resolution transmission electron microscope (HRTEM) as described in Liu *et al*[1]. The chips with nanopores were sealed by silicone o-rings between two polymethylmethacrylate (PMMA) chambers as reservoirs. After mounting, the entire flow cell was flushed with $H_2O$:ethanol (v:v, 1:1) solution and wetted for at least 30 min. This was followed by the injection of 2 M KCl solution buffered with 10mM Tris-HCl and 1mM EDTA at pH 7.0 and $BminPF_6$ (Aldrich-Sigma) to perform current-voltage (IV) characteristics measurements. A pair of chlorinated Ag/AgCl electrodes immersed in two reservoirs and connected to an Axopatch 200B patch clamp amplifier (Molecular Devices, Inc. Sunnyvale, CA) that was used to measure the ionic current as a function of time. Before starting experiments we adjust current offset at zero bias. The device was running at the applied voltage for at least 1 hr to perform blank experiments. DNA samples were diluted in pure $BminPF_6$ by mixing 10 μL of λ-DNA stock solution with $BminPF_6$. DNA samples (pNEB193, plasmid 2.7 k bp, New England; λ-DNA, 48 k bp, New England) were purchased from a commercial supplier, aliquoted and stored at -20 °C before the use. Short homo polymers (Microsynth) and nucleotides (Sigma Aldrich) were purchased in dry form and directly dissolved in RTIL. We use a NI PXI-4461 card for data digitalization and custom-made LabView software for data acquisition. The sampling rate is 100 kHz and a built-in low-pass filter at 10 kHz is used. Data analysis enabling event detection is performed offline using a custom open source Matlab code, named OpenNanopore[4] (http://lben.epfl.ch/page-79460-en.html). For every event, the baseline is recalculated using the average of 100 points before the start of each event. Same criteria have been used for all the compared data. CUSUM algorithm is used to fit the levels inside every event. The current drop is than calculated by subtracting corresponding averaged baseline from the level. Each type of DNA and single nucleotides were translocated in at least two different devices, and representative and reproducible results and analysis are presented.

## COMSOL Modeling

Numerical calculations were performed using the COMSOL 4.2 multiphysics finite-element solver in 3D geometry, imposing a cylindrical symmetry along the axis of the nanopore. We solved the full set of Poisson–Nernst–Planck (PNP) equations, with the boundary conditions at the $MoS_2$ corresponding to an idealized, uncharged membrane impermeable to ions. The PNP set of equations extends Fick's law for the case where the diffusing particles/ions are displaced with respect to the fluid by the electrostatic force. Here we have expressed particle/ion concentrations in terms of mass fractions. In particular, all ion fluxes are modeled by the Nernst-Planck equation

$$\mathbf{J}_i = -D_i \nabla c_i - \frac{F z_i}{RT} D_i c_i \nabla \Phi \qquad (1.0)$$

where $J_i$ and $D_i$, are, respectively, the ion flux vector and diffusion coefficient of species $i$ in the solution, $T$ is the absolute temperature, $\Phi$ is the local potential, $z_i$ is ionic charge and $F$ Faraday's constant. The relationship between the net electric charge of polyelectrolyte and local average electrostatic potential is described by the Poisson's equation.

$$\nabla^2 \Phi(r) = -\frac{\rho(r)}{\varepsilon} \quad (1.1)$$

Both expressions can be rewritten using mass fractions

$$J_i = -\rho D_i^F \nabla \varpi_i + \rho \varpi_i D_i^F \frac{\nabla M_n}{M_n} + D_i^T \frac{\nabla T}{T} + \rho \varpi_i z_i u_{m,i} F \nabla \Phi$$

$$M_n = \left( \sum_i \frac{\varpi_i}{M_i} \right)^{-1} \quad (1.2)$$

where $\varpi_i = \dfrac{m_i}{\sum_i m_i}$ are the mass factors, $\rho$ is the average density, M is the molar mass and $D^F_i$ are diffusion and $D^T_i$ thermal diffusion coefficients and $u$ is the fluid velocity. In the case of 2M aqueous KCl solution (absence of the viscosity gradient), application of a fixed voltage generates the flux of $K^+$ and $Cl^-$ ions that result in the net current that can be easily validated using the well-known analytical expression[5]

$$I = V \left( \left[ \mu_{K^+} + \mu_{Cl^+} \right] n_{KCl} e \right) \left( \frac{4l}{\pi d^2} + \frac{1}{d} \right)^{-1} \quad (1.3)$$

where $V$ is the applied voltage, n is the number density (proportional to concentration) of the ionic species, $e$ is the elementary charge, and $\mu_{K+}$ and $\mu_{CL-}$ are the electrophoretic mobilities of the potassium and chloride ions, respectively. Parameter $d$ represents the pore diameter and $l$ is the membrane thickness. In the case of the viscosity gradient, the set of PNP equations has to be solved for 5 types of diffusing particles/ions subjected to the electrostatic force (4 ions and water molecules). Pore size was fixed to 2.5 nm, 5 nm and 10 nm (see **SI Figure1 (d-f)**). Simulated $MoS_2$ nanopore conductances, for all pore sizes in 2M aqueous KCl solution, viscosity gradient system or pure RT ionic liquid $BminPF_6$ were found to be in a good agreement with the measured values presented in **SI Figure1 (a-c).** In the case of the pure ionic liquid condition and for a fixed voltage, the resulting current originates from the flux of the $Bmim^+$ and $PF_6^-$ ions.

**DNA staining**

DNA was stained with YOYO-1 as described elsewhere[6]. A 1 mM YOYO-1 stock solution (Invitrogen Y3601) was diluted in 10 mM sodium phosphate buffer (1.88 mM $NaH_2PO_4 \cdot H_2O$, 8.13 mM $Na_2HPO_4 \cdot 2H_2O$, pH 7.5) and mixed with λ-DNA with a ratio of 1 YOYO-1 molecule for 5 base pairs.

**Theoretical model**

In **SI Fig. 4a** we schematize the free-energy surface with a well and a barrier to translocation for the case of our viscosity gradient system and for the 2M KCl aqueous solution. In the viscosity gradient system, λ-DNA adopts a random coil configuration with a gyration radius < 240 nm, while in the 2M KCl aqueous solution the corresponding gyration radius is ~570 nm (**SI Fig. 5**). From the schematics, it is obvious that for both systems as long as the applied voltage is lower than the free energy barrier associated with the translocation process, one can expect low probabilities of translocations since they would only be driven by diffusion. On the other hand, increasing the applied voltage reduces the effective barrier and therefore significantly increases the probability of translocations. For the same pore, when working in the 2M KCl aqueous solution, we start to observe translocations at a much lower voltage of 100 mV compared to 200 mV when using the viscosity gradient system **SI Fig. 4b.** This figure shows the comparison between translocation times for pNEB DNA for a wide range of applied voltages and two different electrolyte systems (2M KCl in H$_2$O and the viscosity gradient system). According to Kramer's theory, DNA translocation governed by barrier in both systems obeys a power-law scaling,

$$\tau \sim e^{\left(\frac{\Delta G - qV}{k_B T}\right)} \tag{1.4}$$

where $\tau$ is dwell time, $V$ the applied voltage, $q$ effective charge and $\Delta G$ the height of the free-energy barrier. For both conditions, we observe an exponential dependence that reveals that translocation is voltage-activated, with typical events obtained at different voltages shown in **SI Fig.6.**

**The Stokes drag force in the pores >5 nm**

DNA-pore interactions can slow down DNA translocation in sub-5nm pores [7], while in the larger pores those interactions are negligible. Consequently, in the pores >5nm these interactions should not contribute to the DNA retardation. In the solution, long DNA molecules forms random coils, thus the viscous drag of the whole DNA molecule can be estimated as,

$$F_{drag} = 6\pi\eta v_{DNA} R_g \qquad (1.5)$$

where $R_g$ is the radius of gyration, $\eta$ is the solvent viscosity, and $v_{DNA}$ is the linear velocity of DNA translocation. As the polymer threads through the pore, the center of mass of this sphere moves toward the pore at a velocity:

$$v_{DNA} = \frac{dR_g}{dt} \qquad (1.6)$$

Therefore, the Stokes drag force can be written as,

$$F_{drag} = 6\pi\eta_{IL} R_g \frac{dR_g}{dt} \qquad (1.7)$$

If we assume that DNA translocation velocity is constant, this implies that the force balance between driving force and Stokes drag force is met at all times, *i.e.* from the first monomer translocation to the final monomer translocation.

Then, velocity can be expressed as:

$$v = \frac{R_g}{\tau} \qquad (1.8)$$

where $\tau$ is the translocation time for the entire chain, denoted in experiments as the translocation dwell time. As proposed by Storm *et al.*[8], the principal effect of hydrodynamics is to resist motion with a hydrodynamic drag (Stokes drag) on the DNA coil.

$$F_{Drag} = F_{Driving} \qquad (1.9)$$

In our case with water on both sides of the nanopore,

$$qE = 6\pi\eta R_g \frac{R_g}{\tau} \tag{1.10}$$

we obtain,

$$\tau = \frac{6\pi\eta}{qE} R_g^2 \tag{1.11}$$

Due to the fractal nature of DNA polymers, the equilibrium relation between $R_g$, the radius of gyration of the polymer and DNA length $L_0$ is best described by $R_g=L^\nu$. Then the expression 1.11 for the translocation time of the entire chain can be written as

$$\tau \sim \frac{6\pi\eta}{qE} L_0^{2\nu} \tag{1.12}$$

where $\nu$ is the Flory exponent.

For our viscosity gradient system, we only consider the biggest contribution to the Stokes drag force which originates form the drag of the DNA coil in the *cis* chamber since viscosity of RTIL is much higher than water.

Then,

$$F_{drag} = 6\pi\eta_{IL} R_g^{cis} \frac{dR}{dt} \tag{1.13}$$

where

$$R_g^{cis}(t) = ((N-n)b)^\nu \tag{1.14}$$

where N is the total number of DNA monomers while *n* is the monomer number in the trans chamber and *b* corresponds the monomer length

$$F_{drag} = 6\pi\eta_{RTIL}(N-n)^\nu b^\nu \nu b^\nu (N-n)^{\nu-1} \frac{dn}{dt} \tag{1.15}$$

Introducing the force balance,

$$qE = 6\pi\eta_{RTIL}(N-n)^\nu b^\nu \nu b^\nu (N-n)^{\nu-1}\frac{dn}{dt} \qquad (1.16)$$

$$\int_0^\tau qEdt = \int_0^N 6\pi\eta_{RTIL}(N-n)^\nu b^\nu \nu b^\nu (N-n)^{\nu-1}dn \qquad (1.17)$$

For the viscosity gradient system, $\tau_{RTIL}$ chain translocation time can be written

$$\tau_{RTIL} \sim \frac{3\pi\eta_{IL}L_0^{2\nu}}{qE} \qquad (1.18)$$

At this point we can introduce a retardation factor that allows us to compare between DNA translocation well times obtained aqueous 2M KCl solution in and in the viscosity gradient system

$$r = \frac{\tau_{RTIL}}{\tau_{H_2O}} = \frac{\eta_{RTIL}}{2\eta_{H_2O}} \qquad (1.19)$$

We obtain a retardation factor higher than 140 that is predominantly due to the increase in the viscosity in our viscosity gradient system.

| Translocated molecules | Length (bp, nt) | Supplier |
|---|---|---|
| **Lambda** | 48502 | New England Biolabs |
| **Lambda HindIII** | 125, 564, 2027, 2322, 4361, 6557, 9416, 23130 | New England Biolabs |
| **pNEB 193, plasmid** | 2700 | New England Biolabs |
| **poly A30, T30, G30, C30** | 30 | Microsynth |
| **Single nucleotides** | 1 | Sigma |

**Table S1.** DNA and nucleotides used in this work.

| Device name | Pore size (nm) | Experiments done | Number of events | |
|---|---|---|---|---|
| **100808A** | 4 ± 0.5 | λ-DNA | 80* | Data not shown |
| **100718A** | 22 ± 2 | λ-DNA | 159* | Figure 2 |
| **100803B** | 4.7 ± 1.2 | pNEB plasmid | 1150 | SI Figure 4 and 6 |
| **100829B** | 20 ± 2 | pNEB plasmid | >3000 | Data not shown |
| **100619A** | 9.6 ± 0.8 | λ-DNA and pNEB plasmid | ~1100 | Data not shown |
| **100807B** | 8.3 ± 1 | λ-Hind III digest | 193* | Data not shown |
| **100807A** | 4.2 ± 0.2 | λ-Hind III digest | 242* | Data not shown |
| **100826B** | 2.5 ± 0.5 | λ-Hind III digest | 258* | Data not shown |
| **100619B** | 6.3 ± 0.5 | Poly(A)$_{30}$(T)$_{30}$ | ~1500 | Data not shown |
| **100830B** | 3.7 ± 0.2 | Poly(A)$_{30}$(T)$_{30}$ and Poly(T)$_{100}$ | 1213* | Data not shown |
| **100913A** | 4 ± 1.5 | Poly(A)$_{30}$ | 100* | Data not shown |
| **101003B** | 2.8 ± 0.3 | Homopolymers and single nucleotides | > 10000 | Figure 3 and 4 |
| **101010B** | 2.3 ± 0.1 | dAMP, dTMP and dGMP | >3000 | SI Fig 12 |
| **101014A** | 4 ± 0.2 | dAMP, dTMP and dGMP | 750* | SI Fig 12 |
| **101024B** | 2 ± 0.2 | dAMP and dCMP | 230* | SI Fig 12 |
| **101031B** | 2.2 ± 0.2 | dAMP and dCMP | 1100* | SI Fig 12 |
| **101026A** | 6.9 ± 0.5 | dAMP and dCMP | 160* | Data not shown |
| **101102A** | 13 ± 3 | dAMP and dCMP | 350* | Data not shown |
| **101103A** | 6.5 ± 0.5 | dAMP and dCMP | 3000* | SI Fig 12 |
| **110119A** | 3.3 ± 0.5 | dAMP, dTMP, dCMP and dGMP | >10000 | SI Fig 10,11 and 12 |

**Table S2.** List of MoS$_2$ nanopore devices tested under viscosity gradient conditions. In most of the nanopore devices the collection of events was stopped manually due to the increase in the noise.

# Figures

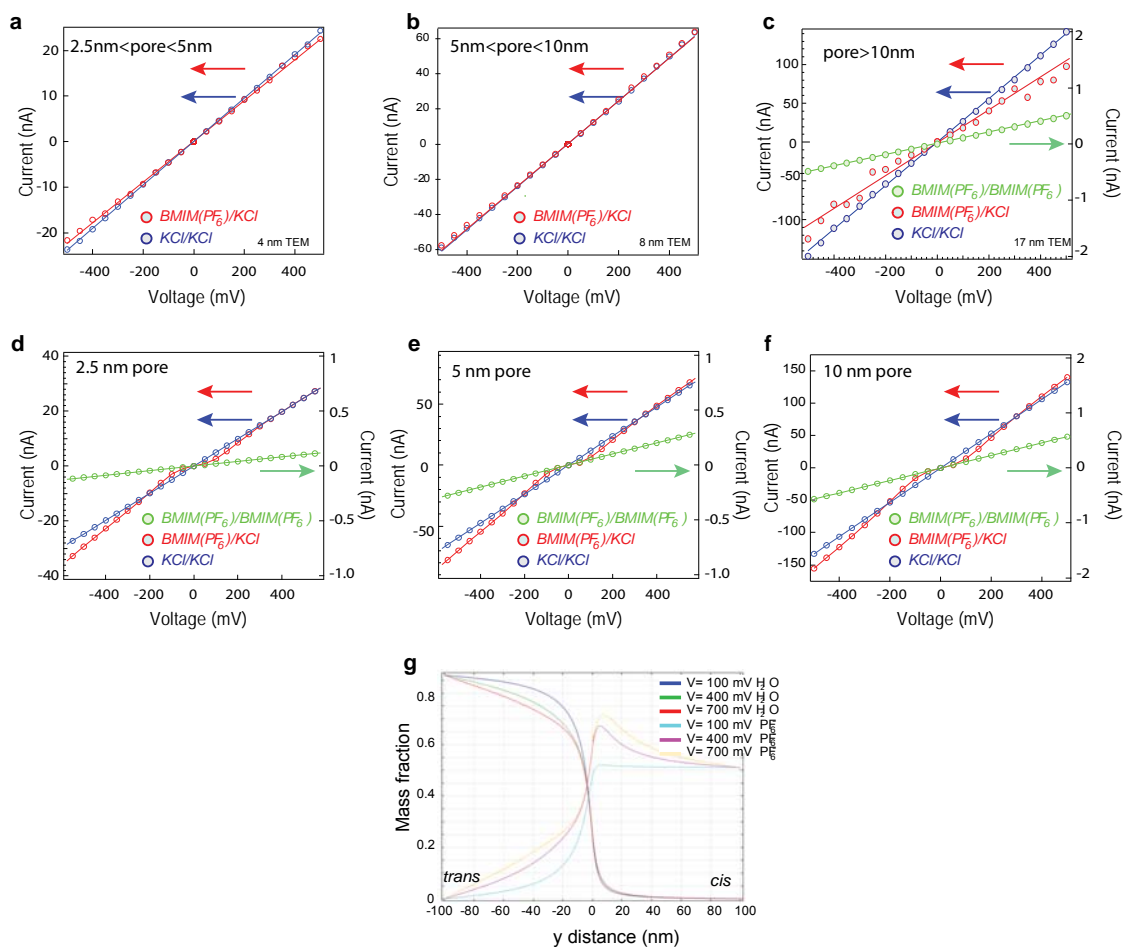

**SI Fig. 1**. **Current-voltage characteristics of MoS$_2$ nanopores and COMSOL simulations of the ionic transport through a MoS$_2$ nanopore.** Measured current–voltage characteristics for viscosity gradient system (red), pure ionic liquid (green) and 2M aqueous KCl solution (blue) (**a**) in a pore smaller than 5 nm (**b**), pore with a diameter between 5 nm and 10 nm (**c**) pore larger than 10 nm. Simulated current–voltage characteristics for a viscosity gradient system (red), pure ionic liquid (green) and 2M aqueous KCl solution (blue) in 2.5 nm pore having a conductance in gradient conditions of ~ 48 nS (**d**), 5 nm pore having conductance in gradient conditions of ~ 120 nS (**e**), and 10 nm pore having conductance in gradient conditions of ~ 280 nS (**f**). (**g**) Mass fraction as a function of distance from the nanopore center (marked as 0) of water, anions (PF$_6^-$ and Cl$^-$), cations (Bmim$^+$ and K$^+$) at different applied voltages.

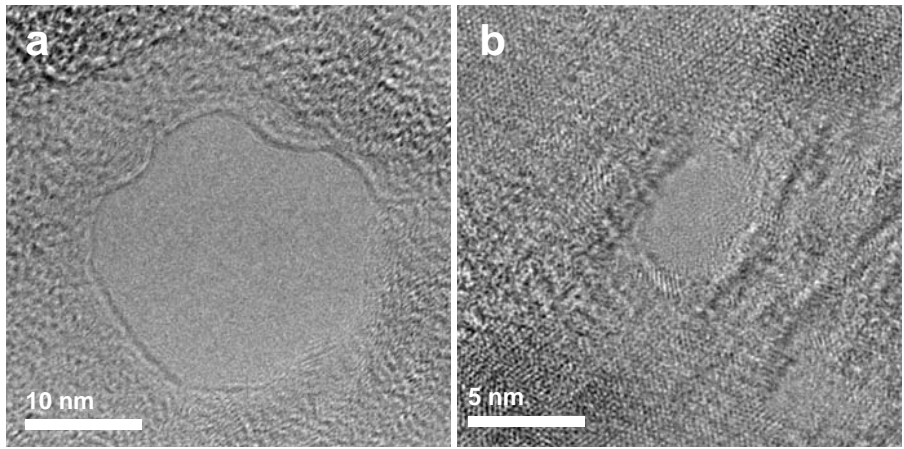

**SI Fig. 2. (a)** High-resolution TEM images of a 22 nm diameter $MoS_2$ nanopore (data shown in Fig. 2) and a 2.8 nm diameter $MoS_2$ nanopore **(b)** (data shown in Fig.3 and 4) drilled using a focused electron beam. The same pore as in (a) has been used in Liu *et.al.* [1].

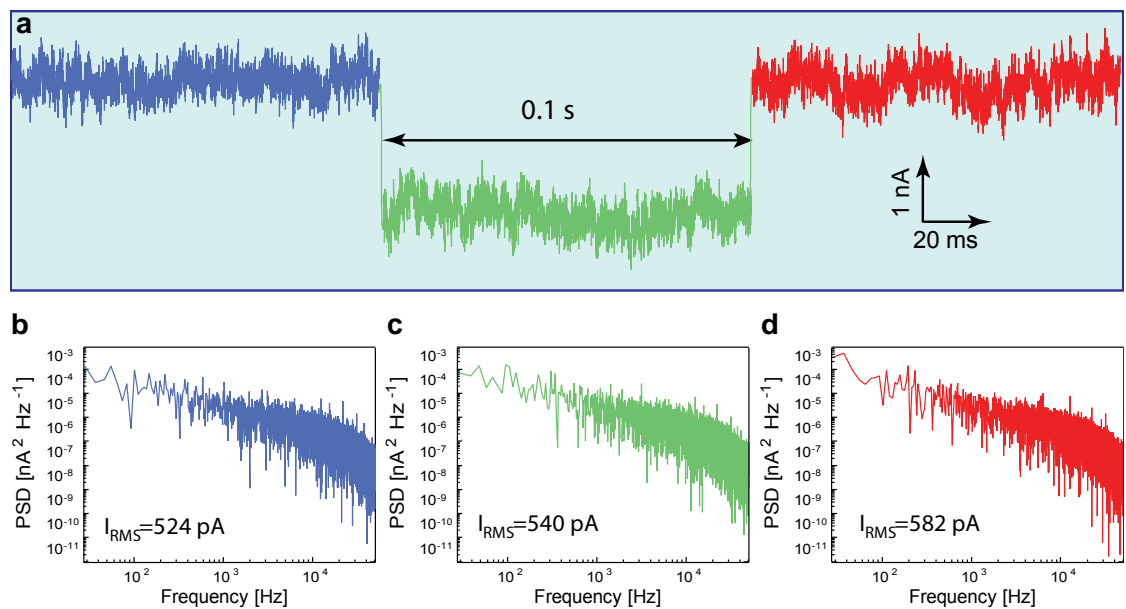

**SI Fig. 3. (a)** An example of a 48.5 kbp λ-dsDNA translocation event in the viscosity gradient system. Current noise power spectra for the trace presented in **(a)** where the noise was calculated using Welch's method from 0.1 seconds of continuous data before DNA translocation (blue **(b)**) during (green **(c)**) and after (red **(d)**) DNA translocation.

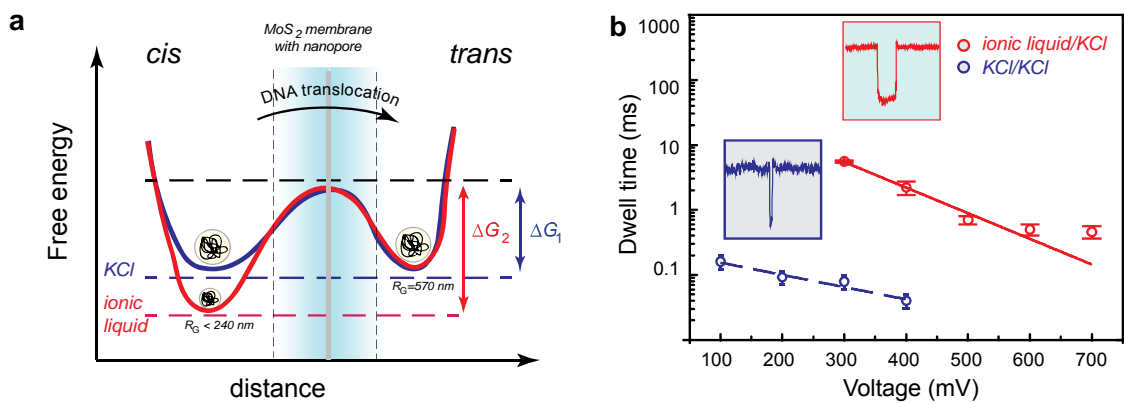

**SI Fig. 4. Single molecule DNA translocation through a nanopore probes the dynamics of Kramer's theory (a)** Schematic representations of single-well free-energy surfaces, for two conditions. The schematics describes the intrinsic (i.e. zero voltage) free-energy surface with a well and a barrier to translocation. In the context of the voltage-driven translocation of individual DNA molecules in a nanopore, the well of the free-energy surface corresponds to the random-coil DNA configuration in a cis chamber with corresponding radius of gyration, while escape over the barrier involves translocation through the nanopore and subsequent adoption of the random-coil conformation. The free energy should include at least two parts, one from the phase transfer as described using L-J equation, another from the entropy part of the DNA coil. Both of these two energy parts give a similar phase as drawn, with the only significant difference being the distance and the free-energy level. **(b)** Dependence of the translocation dwell time on the applied voltage for pNEB DNA in ionic liquid/ KCl solution (red) and in KCl/KCl (blue). For both conditions, we observe an exponential dependence that reveals that translocation is voltage-activated. Blue and red lines are exponential fits to the data.

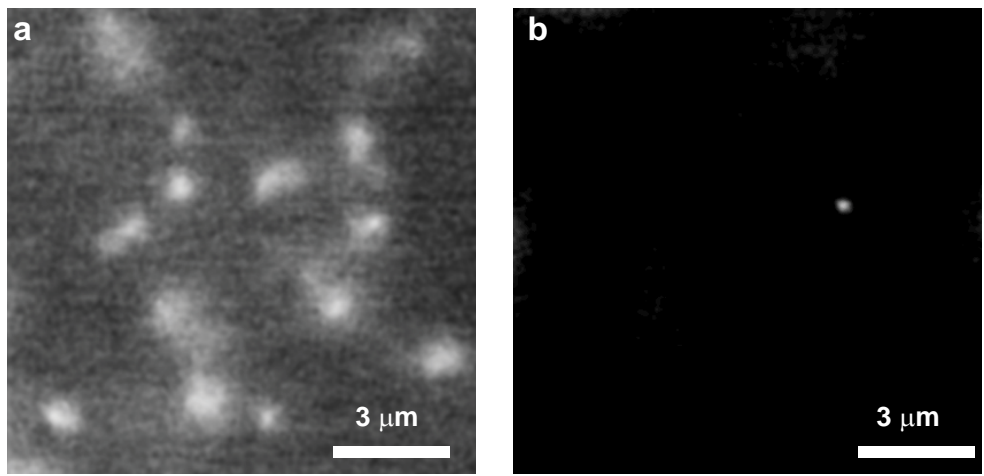

**SI Fig. 5.** Fluorescence images of YOYO-1 labeled λ-DNA extracted from movies used to measure the diffusion coefficient of DNA in water **(a)** and **(b)** in RTIL (BmimPF$_6$). By tracking the locations of individual DNA molecules through a sequence of video frames, one can measure corresponding diffusion coefficients. Same movies were used to extract radius of gyration *Rg* of λ-DNA. In water it was ≈570 nm and less than 240 nm in BmimPF$_6$.

.

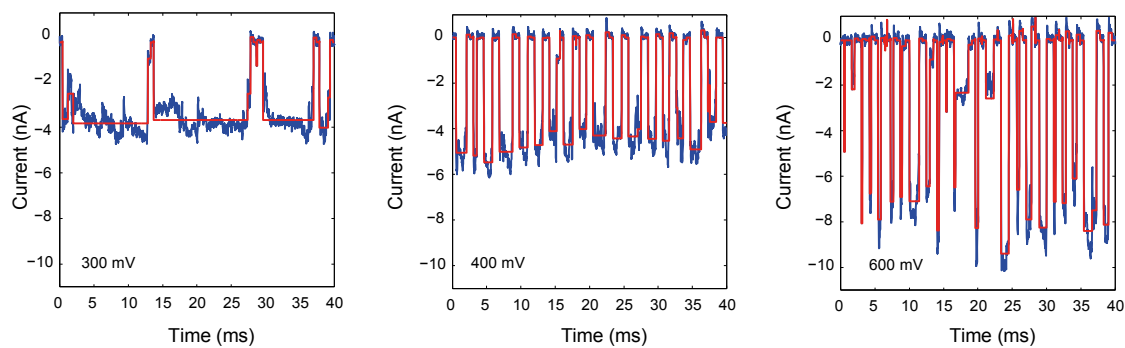

**SI Fig. 6**. Example traces of pNEB translocation traces through a MoS$_2$ pore under ionic liquid/KCl condition with variable voltages (data used for **SI Fig. 4**). The most probable dwell times, from a single-exponential fit, are 5.5 ± 0.2 ms, 2.2 ± 0.5 ms, and 0.5 ± 0.1 ms for 300 mV, 400 mV, and 600 mV, respectively. This also shows a linear relationship between the current signal and applied voltage except at 500 mV (due to baseline fluctuation). We also observed enhanced signal under viscosity gradient conditions.

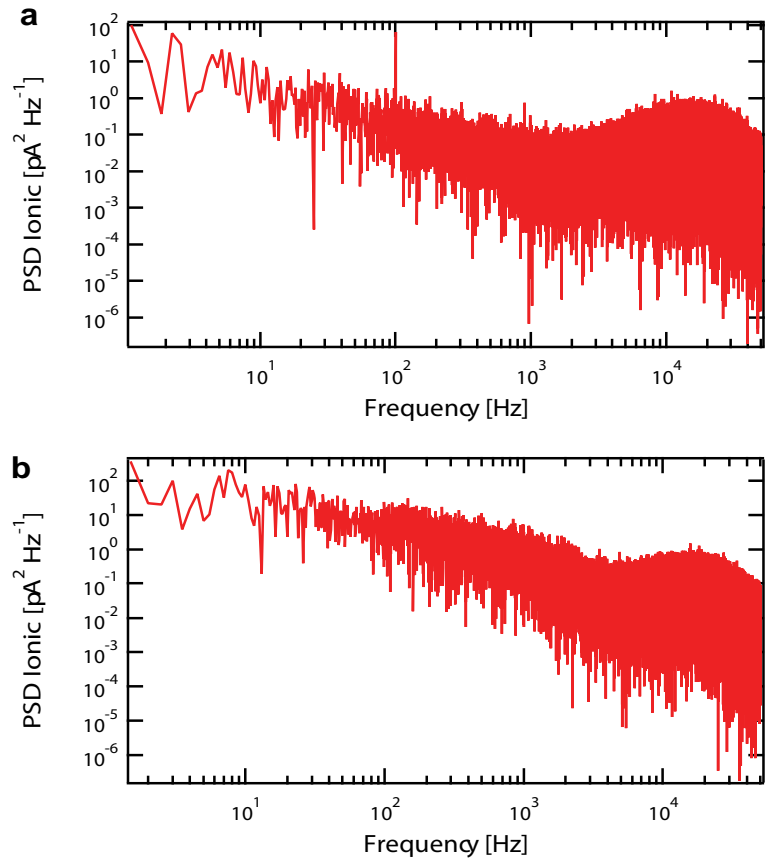

**SI Fig. 7. Current noise power spectra for the 3 nm diameter MoS$_2$ nanopore shown in SI Figure 1b)**. The noise was calculated using Welch's method from 1 second of continuous data before DNA translocation **(a)** at 0 m bias and **(b)** at 200 mV.

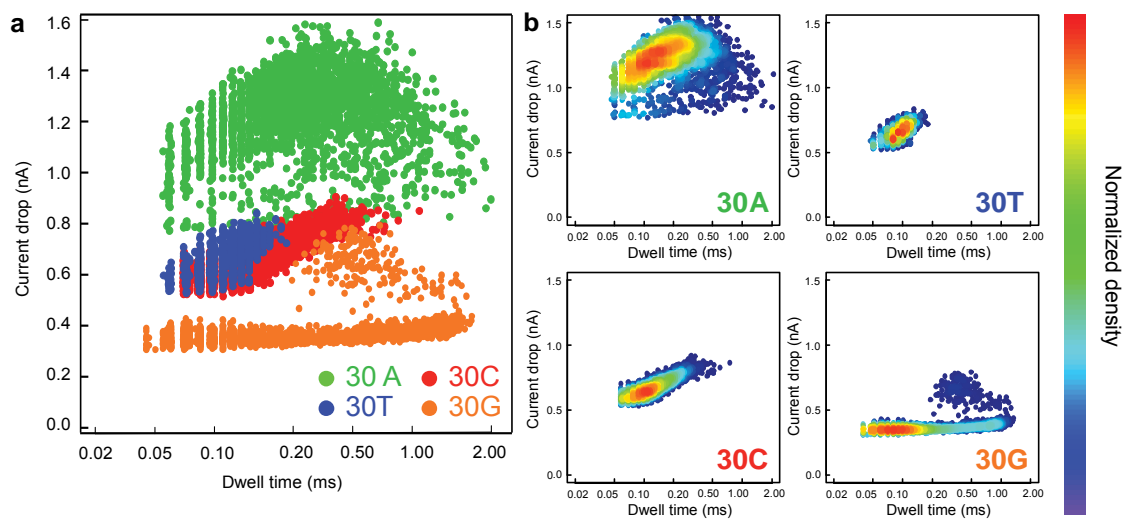

**SI Fig. 8. (a)** Scatter plots of nucleotide translocation, showing distinguished current drops and dwell times for poly A30 (green), poly C30 (red), poly T30 (blue), and poly G30 (orange). b) Normalized histogram of current drops for each kind of the DNA homopolymer. The mean value for poly A30 is 1.25 nA, for poly C30 is 0.65 nA, for poly T30 is 0.7nA and for poly G30 is 0.45 nA. **(b)** Density plots of 30mer oligonucleotides in a MoS$_2$ nanopore; for poly A30, the position of the hot spot is (0.15, 1.25), for poly T30, (0.1, 0.75), for poly C30, (0.12. 0.65) and for poly G30 (0.09, 0.45). The color-map o the right shows the normalized density distribution of events. Data acquired for an experimental condition of pure RTIL in the cis chamber and 100 mM KCl, 25 mM Tris HCl, pH 7.5 in the trans chamber. The bias is +200 mV. The concentration of short DNA homopolymers in RTILs is 0.02 μmol/ml.
20

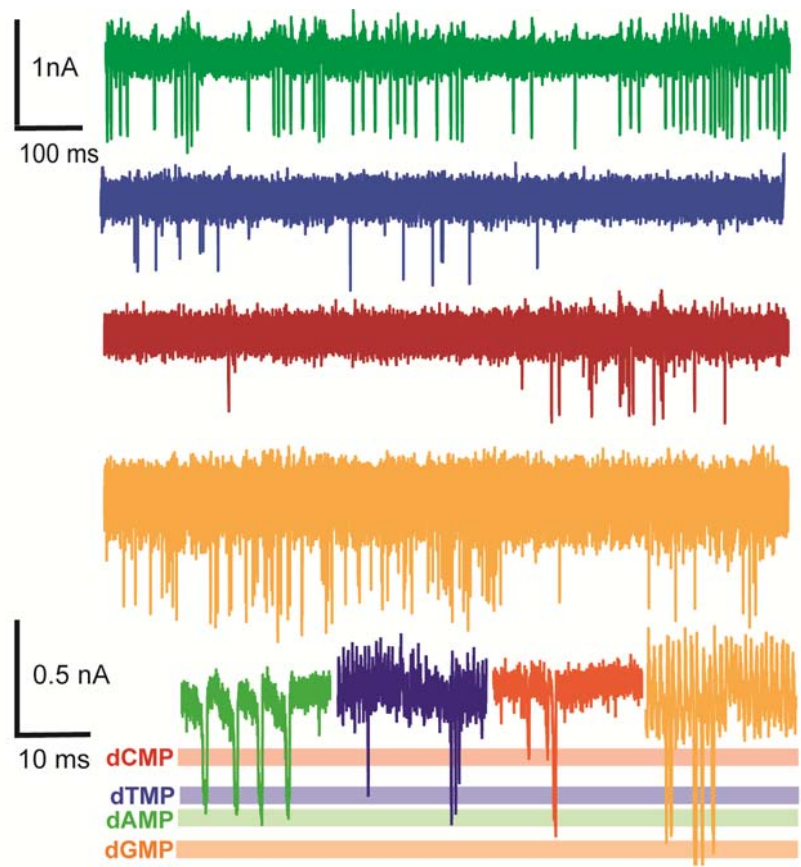

**SI Fig. 9. Differentiation of single DNA nucleotides in the 2.8 nm MoS$_2$ nanopore under ionic liquid/KCl gradient condition**. 0.5 s and 0.1 s translocation signals for each nucleotide dAMP (green), dCMP (red), dTMP (blue), and dGMP (orange).

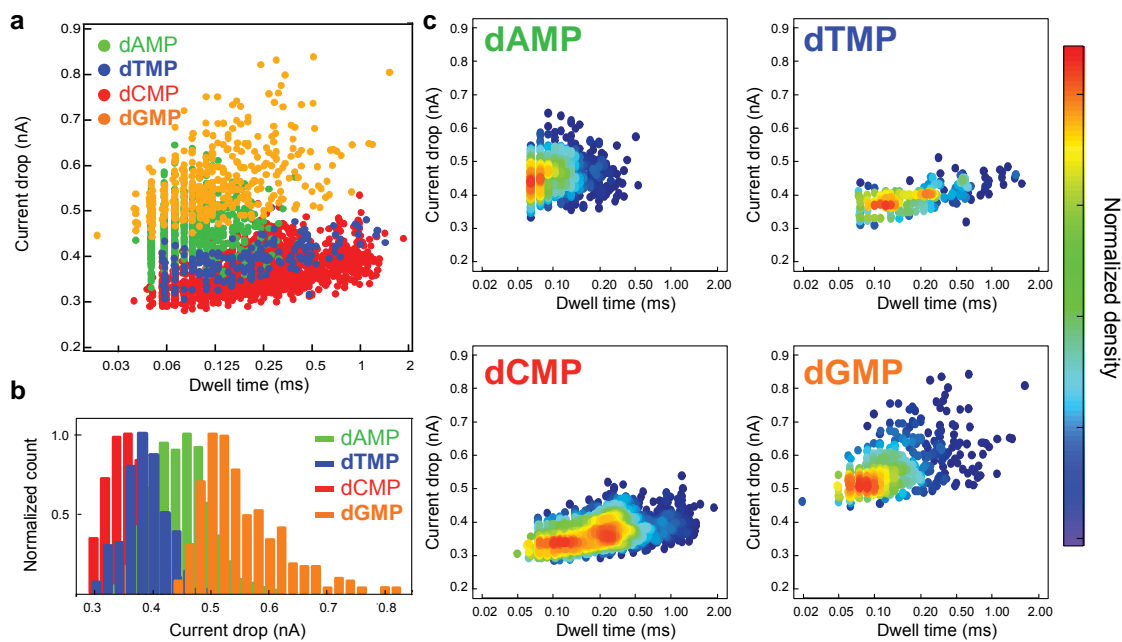

**SI Fig. 10. Identification of single nucleotides in a 3.3 nm MoS$_2$ nanopore**. (**a**) Scatter plots of nucleotide translocation events, showing distinguished current drops and dwell times for dAMP (green), dCMP (red), dTMP (blue), and dGMP (orange). (**b**) Normalized histogram of current drops for dAMP, dTMP, dCMP, dGMP. (**c**) Density plot of single nucleotides in the MoS$_2$ nanopore; for dAMP, the position of the hot spot is (0.07, 0.46), for dTMP, (0.10, 0.40), for dCMP, (0.11. 0.36) and for dGMP (0.08, 0.56). The color-map at the right shows the normalized density distribution of events. It is clear that in the slightly larger pore nucleotide translocation events are faster and have smaller current amplitude drops. However, the trend of current drops for different types of nucleotides remains the same as shown in **Fig 4.** (dGMP>dAMP>dTMP>dCMP). Data acquired for an experimental condition of pure RTIL in the cis chamber and 100 mM KCl, 25 mM Tris HCl, pH 7.5 in the trans chamber. The bias is +200 mV. The nucleotide concentration in RTILs was 5μg/ml.

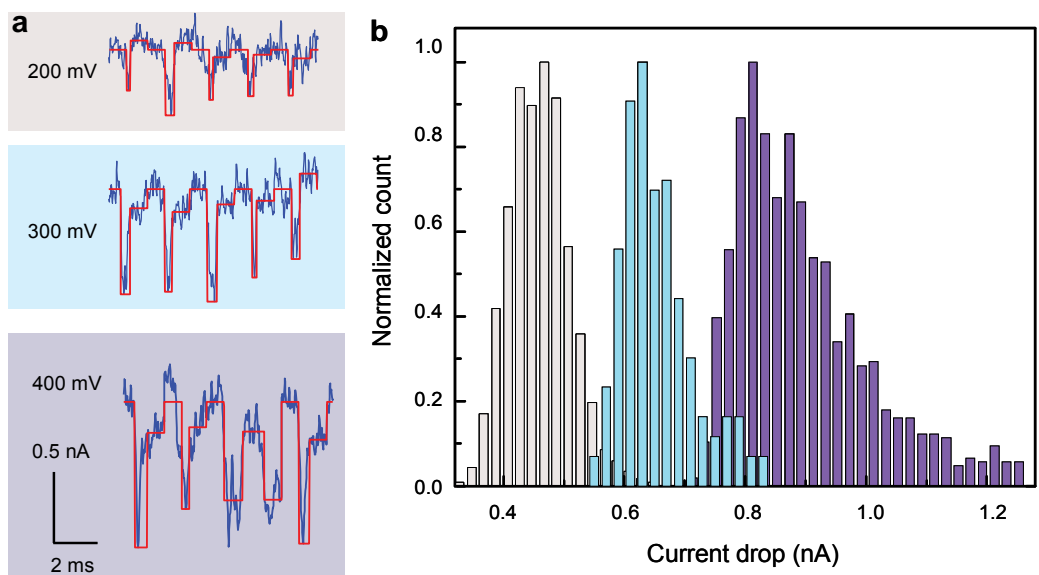

**SI Fig. 11.** Example traces **(a)** and histograms **(b)** of dAMP translocation through a 3.3 nm MoS$_2$ pore in the presence of a viscosity gradient (ionic liquids/KCl), for different voltages (200 mV, 300mV and 400mV). The mean values for current drops are 0.46 nA, 0.65 nA, 0.91 nA, for 200 mV, 300mV, 400mV, respectively.

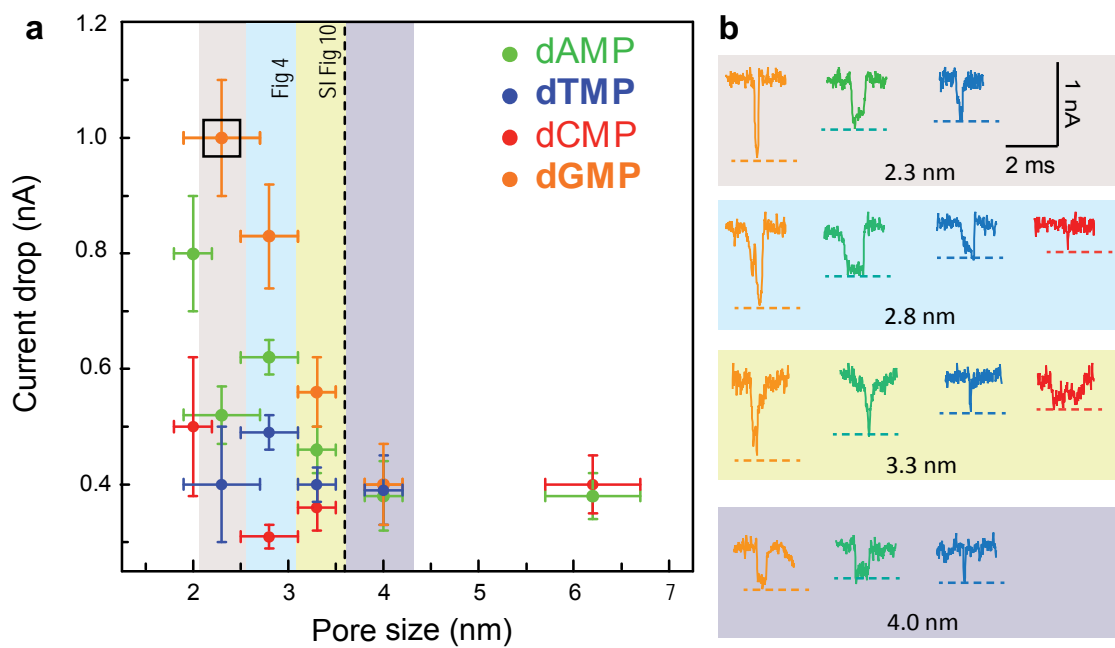

**SI Fig. 12. Pore size dependent differentiation/identification of four nucleotides based on ionic current drops.** (a) Correlation between mean current drops of four nucleotides and pore sizes. Solid circles represent the experimentally determined mean current drops (standard deviation) for dAMP (green), dCMP (red), dTMP (blue), and dGMP (orange), respectively. Errors of pore sizes originate from the asymmetry of electron beam drilled pores. The black dashed line (around 3.6 nm) represents the maximum pore size that still allows differentiating between nucleotides. Nucleotides can be statistically identified within pores smaller than the critical size that is between 3.6 and 4 nm. Black rectangle indicates the data set with highest SNR (~16). (b) Typical events related to four nucleotides (labeled in color) translocating through $MoS_2$ nanopores with different diameters. The levels indicate the mean values for the current drops.